\documentclass[%
 reprint,
 amsmath,amssymb,
 aps,
prb,
]{revtex4-1}

\usepackage{graphicx}
\usepackage{dcolumn}
\usepackage{bm}
\usepackage{amsmath}
\usepackage{amssymb}
\usepackage{txfonts}
\usepackage{mathrsfs}
\usepackage{xcolor}
\usepackage[mathlines]{lineno}

\begin{document}

\title{First-order, continuous, and multicritical Bose-Einstein condensation in Bose mixtures}
\author{Pawel Jakubczyk }
\affiliation{Institute  of Theoretical Physics, Faculty of Physics, University of Warsaw, Pasteura 5, 02-093 Warsaw, Poland}
\author{Krzysztof Myśliwy }
\affiliation{Institute  of Theoretical Physics, Faculty of Physics, University of Warsaw, Pasteura 5, 02-093 Warsaw, Poland}
\author{Marek Napiórkowski }
\affiliation{Institute  of Theoretical Physics, Faculty of Physics, University of Warsaw, Pasteura 5, 02-093 Warsaw, Poland}

\date{\today}

\begin{abstract} 
We address the possibility of realizing Bose-Einstein condensation as a first-order phase transition by admixture of particles of different species. To this aim we perform a comprehensive analysis of phase diagrams of two-component mixtures of bosons at finite temperatures. As a prototype model, we analyze a binary mixture of Bose particles interacting via an infinite-range
(Kac-scaled) two-body potential.  
We obtain a rich phase diagram, where the transition between the normal and Bose-Einstein condensed phases may be either continuous or first-order. The phase diagram hosts lines of triple points, tricritical points, as well as quadruple  points. We address the structure of the phase diagram depending on the relative magnitudes of the inter- and intra-species interaction couplings.
In addition, even for purely repulsive interactions, we identify a first-order liquid-gas type transition between non-condensed phases characterized by different particle concentrations. In the obtained phase diagram, a surface of such first-order transitions terminates with a line of critical points.

\end{abstract}

\maketitle

\section{Introduction} 
Bose-Einstein condensation constitutes a textbook example of a continuous phase transition. The interesting possibility of realizing condensation as a first-order transition was reported in recent studies \cite{Hu_2021, Spada_2023} in setups involving attractive interparticle forces, three-body interactions, and trapping potentials. In this paper we point out that first-order condensation is a phenomenon ubiquitous in Bose mixtures and may be obtained even in rather simple models of homegeneous Bose mixtures with purely repulsive intermolecular interactions. 

Mixtures of quantum fluids have been receiving substantial attention since many years. The early interest motivated primarily by experiments on $^3$He-$^4$He mixtures,\cite{Walters_1956, Graf_1967} became in more recent times boosted by realization of binary mixtures of ultracold alkali atoms.\cite{Ho_1996, Hall_1998, Maddaloni_2000, Delannoy_2001, Modugno_2002} It was quickly noted that such systems may exhibit a variety of ground states,\cite{Esry_1997, Ohberg_1998, Timmermans_1998, Pu_1998, Ao_1998, Shi_2000, Riboli_2002, Altman_2003} which triggered further efforts both on the experiment and theory sides.\cite{Mertes_2007, Bhongale_2008, Papp_2008, Catani_2008, Anderson_2009, Gadway_2010, Hubener_2009, Capograsso_2010, McCarron_2011, Facchi_2011, Lv_2014, Ceccarelli_2015, Lingua_2015, Ceccarelli_2016, Lee_2018, Boudjemaa_2018, Ota_2019, Ota_2020} Despite this, in case of Bose mixtures, 
certain properties of the global phase diagram (in particular at finite temperatures) seem to remain not fully explored. One such aspect concerns the actual order of the transition between the normal and Bose-Einstein-condensed phases depending on the thermodynamic parameters. Recent results of numerical simulations (see Ref. \onlinecite{Spada_2023}) suggested that in the case of mixtures with interactions involving an attractive component (see \onlinecite{Petrov_2015, Semeghini_2018, Naidon_2021}),
the transition between the normal phase and the phase hosting a Bose-Einstein condensate (BEC) may actually be of first-order. Another recent study reports condensation as a first-order transition \cite{Hu_2021} even for single-component systems, but invokes attractive two-body and repulsive three-body interactions. In fact, suggestions concerning the possible first-order condensation  were made earlier \cite{Schaeybroeck_2013} also in reference to the simple Bose mixtures with purely repulsive microscopic interactions. We are however not aware of any systematic study of this issue.  


In the present paper we address the phase diagram of the two-component Bose mixture, employing the exactly soluble imperfect Bose gas model with purely repulsive interactions and, for sufficiently strong interspecies repulsion, demonstrate realization of Bose-Einstein condensation as a first-order transition. In addition, we identify another transition between two non-condensed phases characterized by different concentrations of the mixture constituents. We are not aware of this transition being discussed in previous studies of this system.  

For the one-component case, the model employed by us was first discussed in Ref. \onlinecite{Davies_1972} (see also Refs.~\onlinecite{Buffet_1983, Berg_1984, Lewis_1985, Smedt_1986, Zagrebnov_2001}). Its physical content is clarified by the Kac scaling procedure, see e.g. Ref. \onlinecite{Hemmer_1976}, where a realistic two-body interaction potential $v(\vec{x})$ is promoted to the form 
\begin{equation} 
\label{Kac_sc} 
v(\vec{x}) \rightarrow v_\gamma(\vec{x})=\gamma^d v(\gamma \vec{x})\;, 
\end{equation}
which depends on a positive parameter $\gamma$. Here $d$ denotes the spatial dimensionality of the system and $\int_{\mathbb{R}^d} d\vec{x}\; v_\gamma(\vec{x})=a>0$  is clearly independent of $\gamma$. The imperfect Bose gas model corresponds to taking the limit $\gamma\to 0^+$, where the interaction becomes very weak and long range. In this limit one finds the usual two-body interaction part of the Hamiltonian $\hat{H}_{int}$
\begin{equation} 
\label{H_inter} 
\hat{H}_{int}=\frac{1}{2V}\sum_{\vec{k}, \vec{k'}, \vec{q}} v_{\vec{q}} a^{\dagger}_{\vec{k}+\vec{q}}  a^{\dagger}_{\vec{k'}-\vec{q}} a_{\vec{k}'} a_{\vec{k}}\;, 
\end{equation}
where $v_{\vec{q}}$ is the Fourier transform of $v(\vec{x})$ and $V$ denotes the system volume, to be simplified as follows
\begin{equation} 
\hat{H}_{int} \longrightarrow \hat{H}_{int}^{IBG} = \frac{v_{\vec{0}}}{2V} \sum_{\vec{k}, \vec{k}'} a^{\dagger}_{\vec{k}}  a^{\dagger}_{\vec{k'}} a_{\vec{k}'} a_{\vec{k}} =\frac{a}{2V}\hat{N}(\hat{N}-1)\;,  
\end{equation}
where $\hat{N}=\sum_{\vec{k}}a^{\dagger}_{\vec{k}} a_{\vec{k}}$ denotes the total particle number operator. In the thermodynamic limit, on which we here focus, the last term may be simplified by replacing $(\hat{N}-1)\rightarrow \hat{N}$. We consider spin-zero bosons. 

No approximation is involved in the above transformation, which amounts to taking the limit $\gamma\to 0^+$ in Eq.~(\ref{H_inter}). The resulting model corresponds to 
infinitely weak and long-ranged interactions and, as such, can be solved exactly by a saddle-point approximation in the thermodynamic limit, as we demonstrate below (see Sec.~II).  

The single component imperfect Bose gas in the continuum is defined by 
\begin{equation}
\hat{H}_{IBG}= \sum_{\vec{k}}\frac{\hbar^2 \vec{k}^2}{2m} a^{\dagger}_{\vec{k}} a_{\vec{k}} + \hat{H}_{int}^{IBG} 
\label{IBGH}
\end{equation}
and its phase diagram as well as critical behavior was fully clarified in Refs.~\onlinecite{Napiorkowski_2011, Napiorkowski_2013, Diehl_2017,  Jakubczyk_2018}. Despite some similarity to the perfect Bose gas, its behavior is significantly closer to realistic, interacting systems. In particular (in contrast to the perfect Bose gas): $(i)$ it exhibits superstability \cite{Lewis_1984} such that its descriptions using  distinct Gibbs ensembles are fully equivalent ; $(ii)$ its thermodynamics is defined both for negative and positive values of the chemical potential; $(iii)$ the transition to the BEC phase is of second order and characterized by non-classical critical exponents. More specifically, it falls into the universality class of the spherical (Berlin-Kac) model,\cite{Berlin_1952} corresponding also to the limit $N\to \infty$ of $O(N)$-symmetric models.\cite{Stanley_1968, Moshe_2003} For temperature $T\to 0$ the model displays a quantum critical point characterized by a dynamical exponent $z=2$,\cite{Jakubczyk_2013, Jakubczyk_2016, Frerot_2022} which can be accessed by varying the chemical potential between positive and negative values. 

The paper is structured as follows: in Sec.~II we introduce a simple generalization of the model defined in Eq.~(\ref{IBGH}) to account for Bose mixtures and present the analytical part of its solution, which becomes exact in the thermodynamic limit. In Sec.~III we outline the  procedure leading to the practical determination of the phase diagram. In Sec.~IV we analyze the asymptotic behavior of the system at low concentration of one of the mixture constituents. In Sec.~V we present the major results concerning the phase diagram depending on relative magnitudes of the interaction couplings. 
In Sec.~VI we focus on the normal phase in the regime of relatively strong interspecies interactions and analyze the liquid-gas-type transition occurring therein between two non-condensed phases involving different particle concentrations. 
We summarize the paper in Sec.~VII and some technical details are given in the appendices.

\section{The imperfect Bose mixture} 
We propose a simple generalization of the imperfect Bose gas model described above to account for binary Bose mixtures. The Hamiltonian is defined as: 
\begin{equation} 
\label{imp_B_mix}
\hat{H}=\sum_{\vec{k},i}\epsilon_{\vec{k},i}\hat{n}_{\vec{k},i}+\sum_{i, j}\frac{a_{i,j}}{2V}\hat{N}_i \hat{N}_{j}\;.
\end{equation}
Here $i,j\in\{1,2\}$, $a_{1,2}=a_{2,1}>0$, which will be denoted as $a_{12}$ is the interspecies coupling, while the intraspecies couplings 
$a_{i,i}>0$ are assumed positive and will be denoted as $a_i$. The dispersion takes the standard form $\epsilon_{\vec{k},i}=\hbar^2\vec{k}^2/(2m_i)$. The system is subject to periodic boundary conditions. We use the grand-canonical ensemble, where the grand-canonical free energy density $\omega(T,\mu_1, \mu_2)$ is obtained from the grand canonical partition function $\Xi(T,V,\mu_1,\mu_2)$ in the thermodynamic limit: 
\begin{equation}
\omega(T,\mu_1, \mu_2)=-\beta^{-1}\lim_{V\to\infty}\frac{1}{V}\log\Xi(T,V,\mu_1,\mu_2)    
\end{equation}
and 
\begin{equation} 
\Xi(T, V, \mu_1, \mu_2)={\textrm{Tr}}\, e^{-\beta (\hat{H}-\mu_1 \hat{N}_1 -\mu_2 \hat{N}_2)}\;. 
\end{equation}
Here $\beta^{-1}= k_B T$ and $\mu_1$, $\mu_2$ are the chemical potentials of the two mixture constituents. By a sequence of exact transformations described in the Appendix 1, the partition function may be cast in the following form  
\begin{equation} 
\label{Xixi}
\Xi(T,V,\mu_1,\mu_2)=-\frac{\beta V}{2\pi\sqrt{a_1' a_2'}}\int_{\alpha_1-i\infty}^{{\alpha_1+i\infty}}dt_1 \int_{\alpha_2-i\infty}^{{\alpha_2+i\infty}}dt_2 e^{-V\Phi(t_1,t_2)}\;, 
\end{equation}
valid for  $a_1 a_2-a_{12}^2>0$. 
Here
\begin{align}
\Phi(t_1, t_2)=& -\sum_{i=1}^2\frac{\beta}{2a_i'}(t_i-\mu_i')^2-\frac{1}{\lambda_1^3}g_{5/2}(e^{\beta t_1})-\frac{1}{\lambda_2^3}g_{5/2}(e^{\beta (\frac{a_{12}}{a_1}t_1+t_2)})  \\ 
&+\frac{1}{V}\log (1-e^{\beta t_1}) +\frac{1}{V}\log (1-e^{\beta (\frac{a_{12}}{a_1}t_1+t_2)})\;, \nonumber
\end{align} 
 $a_1'=a_1$, $a_2'=a_2(1-\frac{a_{12}^2}{a_1 a_2})$, $\mu_1'=\mu_1$, $\mu_2'=\mu_2-\frac{a_{12}}{a_1}\mu_1$, and $\lambda_i=h/\sqrt{2\pi m_i k_B T}$ are the thermal de Broglie lengths. The Bose functions $g_\alpha (x)$ are defined as:
 \begin{equation} 
 g_\alpha (x)=  \sum_{k=1}^\infty \frac{x^k}{k^\alpha}\;.
 \end{equation}

 An analogous expression involving the same function $\Phi(t_1, t_2)$, but different integration contours in the complex planes can be derived for the complementary range $a_1 a_2-a_{12}^2<0$ - see Appendix 1. 
The quantities $\alpha_1$ and $\alpha_2$ are arbitrary real parameters. From the structure of the above expressions it follows that the integrals defining $\Xi(T,V,\mu_1,\mu_2)$ may be evaluated using the saddle-point approximation, which becomes exact in the thermodynamic limit due to the presence of the $V$-factor in the exponential on the right-hand side of Eq.~(\ref{Xixi}). The stationarity condition reads: 
\begin{equation} 
\label{stat_cond}
\frac{\partial \Phi}{\partial t_1}=0 \;,\;\;\;\;  \frac{\partial \Phi}{\partial t_2}=0   
\end{equation}
and we denote the solution as $(\bar{t_1}, \bar{t_2})$. The grand-canonical free energy density becomes: 
\begin{equation} 
\omega (T,\mu_1, \mu_2)=-\beta^{-1}V^{-1}\log \Xi \longrightarrow \beta^{-1}\Phi(\bar{t_1}, \bar{t_2})\;. 
\end{equation}
The densities $n_i$ $(i\in\{1,2\})$ follow from 
\begin{equation}
n_i(T,\mu_1,\mu_2)=-\frac{\partial\omega}{\partial\mu_i}=-\beta^{-1}\frac{\partial\Phi(\bar{t_1}, \bar{t_2})}{\partial \mu_i}\;.    
\end{equation}
This, together with the stationarity condition of Eq.~(\ref{stat_cond}), leads to the following simple relations: 
\begin{align} 
    n_1 =& -\frac{1}{a_1'}(\bar{t_1}-\mu_1')+\frac{1}{a_2'}(\bar{t_2}-\mu_2')\frac{a_{12}}{a_1} \nonumber \\
    n_2 =& -\frac{1}{a_2'}(\bar{t_2}-\mu_2')  \;.
\end{align} 
Using the above relations one may eliminate $\bar{t_i}$ from the saddle-point equations, which leads to: 
\begin{align}
n_1=&\frac{1}{\lambda_1^3}g_{3/2}(e^{\beta(\mu_1-a_1 n_1- a_{12}n_2)})+\frac{1}{V} \frac{e^{\beta(\mu_1-a_1 n_1- a_{12} n_2)}}{1-e^{\beta(\mu_1-a_1 n_1- a_{12}n_2)}}    \label{n1eq}\\ 
n_2=&\frac{1}{\lambda_2^3}g_{3/2}(e^{\beta(\mu_2-a_2 n_2- a_{12}n_1)})+\frac{1}{V} \frac{e^{\beta(\mu_2-a_2 n_2- a_{12} n_1)}}{1-e^{\beta(\mu_2-a_2 n_2- a_{12}n_1)}} \label{n2eq}\;.
\end{align}
In  the thermodynamic limit, on which we here focus,  the terms $\sim\frac{1}{V}$ in the above equations contribute to the condensate densities of the two mixture constituents. This can be shown by analyzing the densities of particles with momentum $\vec{k}=0$. These quantities  
 will from now on be denoted as $n_1^{(0)}$ and $n_2^{(0)}$, respectively.  We observe that in presence of condensation we have $n_{i}^{(0)} = n_{i} - n_{i,c}$, where $n_{i,c} \equiv \zeta(3/2)\,\lambda_{i}^{-3}$ and $\zeta(x)$ denotes the Riemann zeta function. Note that in particular $\zeta(3/2)=g_{3/2}(1)$. 
 
The above equations for the densities were derived from the model defined in Eq.~(\ref{imp_B_mix}) and in Sec.~I by a sequence of exact transformations. However, their structure reveals affinity to the self-consistent Hartree-Fock (H-F) treatment of dilute Bose gases interacting via short-ranged interactions (see e.g. Ref.~\onlinecite{Schaeybroeck_2013}). The expressions for the thermal densities given by Eq.~(\ref{n1eq}, \ref{n2eq}) above are equivalent to those resulting from the H-F approximation upon identifying $a_i\rightarrow 8\pi\hbar^2 a_i^s/m_i$ and $a_{12}\rightarrow 2\pi\hbar^2 a_{12}^s(m_1^{-1}+m_2^{-1}) $, where $a_i^s$ and $a_{12}^s$ are the corresponding scattering lengths.

By expressing $\bar{t_i}$ via the densities one may also rewrite the quantity $\Phi$ in terms of $n_1$ and $n_2$. We find: 
\begin{align} 
&\Phi(n_1,n_2)= -\frac{\beta}{2}\left[a_1 n_1^2 +a_2n_2^2 +2a_{12} n_1 n_2\right] \nonumber \\
              & -  \frac{1}{\lambda_1^3}g_{5/2}\left(e^{\beta(\mu_1-a_1 n_1- a_{12}n_2)}\right) -\frac{1}{\lambda_2^3}g_{5/2}\left(e^{\beta(\mu_2-a_2 n_2- a_{12}n_1)}\right) \nonumber \\
              & +\frac{1}{V}\log\left(1-e^{\beta(\mu_1-a_1 n_1- a_{12}n_2)} \right) \label{Phieq} \nonumber\\
              & +\frac{1}{V}\log\left(1-e^{\beta(\mu_2-a_2 n_2- a_{12}n_1)} \right)\;.              
\end{align}
The last two terms in the above expression always vanish in the thermodynamic limit (contrary to their counterparts in Eqs (\ref{n1eq},\ref{n2eq}) which contribute to the condensate densities). The reason for this is due to the presence of the logarithm and is in full analogy to the noninteracting case. 
Equations (\ref{n1eq}-\ref{Phieq}) constitute the starting point for the analysis leading to the determination of the system's phase diagram. We emphasize at this point that the quantity $\beta^{-1}\Phi(n_1,n_2)$ has the physical meaning of the (grand-canonical) free energy only when evaluated at the physical values of $(n_1,n_2)$ obtained at  saddle points and should not be understood as any free energy functional. In particular  the above analysis [see Eq.~(\ref{Xixi})] necessarily requires considering $\Phi(n_1,n_2)$ in the complex domain. One may also note that when the analysis  is implemented in the simpler and well-studied case of a single-component system, an analogous quantity [$\Phi(n)$], when viewed as a function of a complex variable, features a saddle-point at the equilibrium density $n=\bar{n}$ located on the real axis. However, if $\Phi(n)$ is considered as only a function of a real variable, $\bar{n}$ is easily shown to be the maximum of $\Phi(n)$.   
We also observe, that the expression for $\Phi(n_1,n_2)$ bears similarity, but is not equivalent to the H-F expression for the density-dependent grand canonical potential (see e.g. Ref.~\onlinecite{Schaeybroeck_2013}). This is in contrast to the expressions for the thermal densities as implied by Eq.~(\ref{n1eq}, \ref{n2eq}).  

The procedure implemented by us can be summarized as follows: for fixed values of $\{a_1, a_2, a_{12}, \lambda_1, \lambda_2, \mu_1, \mu_2, \beta \}$ we solve Eqs. (\ref{n1eq}, \ref{n2eq}) and determine the densities $n_1$, $n_2$. As turns out, for many choices of the system parameters more than one solution for $n_1$, $n_2$ is identified. In such cases, we pick the one corresponding to the lower value of $\Phi (n_1,n_2)$. The thermodynamic state of the system is then determined depending on whether $n_i^{(0)}=0$ or $n_i^{(0)}>0$. In addition, in the normal (non BEC) state we find two phases characterized by distinct density composition (see Sec. VI). First-order transitions are identified as discountinuities of the densities as functions of the system parameters. We give more details of the procedure together with the results in Secs.~III and IV below. 

\section{The solution procedure}
It follows from Eq. (\ref{n1eq})  that absence of condensation of component 1 requires that 
\begin{equation} 
\label{nocondcond}
\mu_1-a_1 n_1-a_{12} n_2<0\;
\end{equation}
which corresponds to $0<n_1<n_{1,c}$ On the other, hand condensation of type-1 particles takes place for 
\begin{equation} 
\label{condcond}
\mu_1-a_1 n_{1,c}-a_{12} n_2=0\;.    
\end{equation}
Analogous conditions hold for the absence/presence of condensate of type-2 particles. Assuming absence of type-1 condensation, i.e. $n_{1}^{(0)} = 0$ (which is then consistently checked), we may determine $n_2$ from Eq.~(\ref{n1eq}) 
\begin{equation}
n_2=-\frac{1}{a_{12}}\left\{\beta^{-1}\log \left[g_{3/2}^{-1}\left(\lambda_1^3 n_1\right)\right]-\mu_1+a_1 n_1\right\}    
\end{equation} 
and insert this into Eq.~(\ref{Phieq}), which yields $\Phi (n_1,n_2(n_1))$.  We subsequently analyze $\Phi (n_1,n_2(n_1))$ as a function of $n_1$ for various choiced of the system parameters. We then check consistency with Eq.~(\ref{nocondcond}) and the analogous condition for the absence of type-2 condensate.  

A distinct case occurs if type-1 particles condense. In such situation we determine $n_2$ from Eq.~(\ref{condcond}) and plug into Eq.~(\ref{Phieq}) to obtain $\Phi (n_1,n_2(n_1))$ (which obviously differs from the previous case). We subsequently find stationary points of the resulting function and check their consistency with the assumptions made [i.e. $n_1>n_{1,c}$ and $0<n_2<n_{2,c}$]. In the same manner we treat the case involving condensation of type-2, but not type-1 particles. Finally we analyze the possibility of obtaining a state hosting condensates of both type-1 and type-2 particles. In this case both $n_1$ and $n_2$ are  obtained as simple, linear functions of $\mu_1$ and $\mu_2$ from Eq.~(\ref{condcond}) and the analogous condition for condensation of type-2 particles. By plugging these into 
Eq.~(\ref{Phieq}) one straightforwardly obtains the expression for $\Phi$ valid for $n_1> n_{1,c}$, $n_2> n_{2,c}$ By comparing the values of $\Phi$ corresponding to all solutions, we project out the phase diagrams as described in the following sections. 

\section{Asymptotic regimes} 
\begin{figure}
\includegraphics[width=8.5cm]{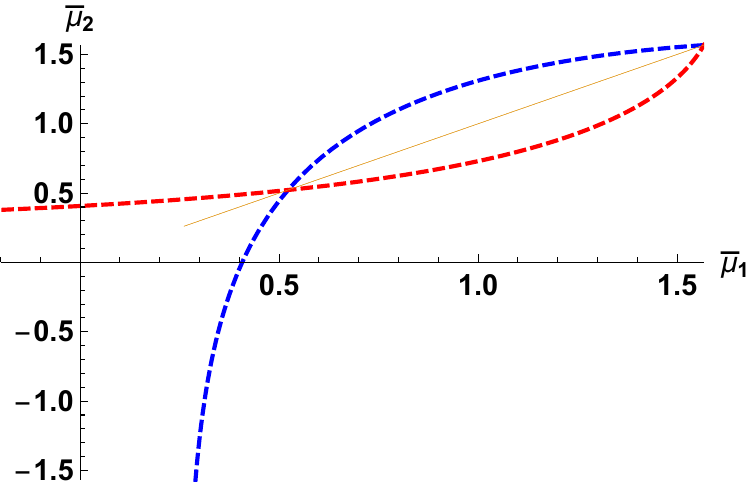}
\caption{The putative critical lines for Bose-Einstein condensation of particles type 1 (blue dashed line) and particles type 2 (red dashed line) plotted in the $(\overline{\mu}_{1}, \overline{\mu}_{2})$-plane at fixed temperature; see Appendix 2 for the definitions of the dimensionless variables implemented in the plot.  The thin yellow line marks $\overline{\mu}_2=\overline{\mu}_1$. The asymptotes are situated at $\overline{\mu}_i=\overline{a}_i\zeta(3/2)$ and the higher point of intersection is located at $\overline{\mu}_1=\left(\overline{a}_1 + \overline{a}_{12} \kappa\right)\zeta(3/2)$. The true transition lines are described by these solutions only for $\overline{\mu}_1$ or $\overline{\mu}_2$ sufficiently  low (see the main text). The plot parameters are $\overline{a}_1 = \overline{a}_2=0.1$, $\overline{a}_{12}=0.5$, $\kappa = 1$.} 
\label{Fig_1}
\end{figure}

We now analyze the system in the asymptotic regime, where concentration of one of the constituents of the mixture, say type-2 particles, is sufficiently low and has minor impact on the behavior of type-1 particles. This corresponds to $\mu_2$ negative and sufficiently large in value. We investigate the Bose-Einstein condensation for type-1 particles, assuming that type-2 particles do not condense and the densities evolve continuously as functions of the control parameters ($T$, $\mu_1$, $\mu_2$). At the BEC transition for type-1 particles we have: 
\begin{align}
n_1=n_{1,c}\,, \;\;
\mu_1=  a_1 n_1+a_{12} n_2\;.
\end{align}
It follows that $n_2=\frac{1}{a_{12}}\left(\mu_1-a_1 n_{1,c}\right)$. When plugging this into Eq.~(\ref{n2eq}), we obtain 
\begin{align} 
 \mu_2^{(1,c)} =&\frac{a_2}{a_{12}}\mu_1-\frac{n_{1,c}}{a_{12}}(a_1 a_2-a_{12}^2) \nonumber \\+&\beta^{-1}\log\left\{g_{3/2}^{-1}\left[\frac{\lambda_2^{3}}{a_{12}}\left(\mu_1-a_1 n_{1,c}\right)\right]\right\} \label{mu2ceq}\;. 
 \end{align}
The above formula describes the putative location of the critical surface, where type-1 particles undergo condensation. Its derivation only requires the self-consistent equations  for the densities [Eq.~(\ref{n1eq}, \ref{n2eq})] and may also be recovered from the H-F treatment of the dilute Bose gas.\cite{Schaeybroeck_2013} A representative plot of its projection on the ($\mu_1, \mu_2$) plane is shown in Fig.~\ref{Fig_1}. 
Note that existence of $\mu_2^{(1,c)}$ requires that 
\begin{equation}
0\leq \frac{\lambda_2^3}{a_{12}}\left(\mu_1- a_1\,n_{1,c}\right)\leq \zeta(3/2)\;,
\end{equation}
which may also be rewritten as $0\leq n_2\leq n_{2,c}$. We obtain:
\begin{equation}
    \mu_1 \in \left[ a_1\,n_{1,c},\,\,a_1\,n_{1,c} + a_{12}\,n_{2,c}\right]\;.
\end{equation}
The lower bound $a_1\,n_{1,c}$ marks position of the vertical asymptote of the projection of the critical line on the $(\mu_1, \mu_2)$ plane - compare Fig. \ref{Fig_1}. 
By interchanging the roles of $n_1$ and $n_2$ we may obviously also obtain an  expression for the critical chemical potential for condensation of type 2 particles (assuming this time that particles of type 1 do not condense).  

At the beginning of this section we introduced the physical assumption that the present analysis is restricted to the regime of small concentrations of one of the mixture constituents. On the other hand, the above derivation of $\mu_2^{(1,c)}$ and analogously of $\mu_2^{(2,c)}$ is not based on this assumption in any way. Indeed, the identified solutions to Eq.~(\ref{n1eq}, \ref{n2eq}) are valid for any $(\mu_1, \mu_2)$ as long as they make mathematical sense. As we demonstrate below, these analytical solutions correspond to a minimum of $\Phi(n_1, n_2(n_1))$ exclusively for $\mu_2$ (or $\mu_1$) sufficiently low. Nonetheless, they correctly describe the second-order transition in a substantial range of the phase diagram. In the complementary case, they instead fall at the maximum of $\Phi (n_1, n_2(n_1))$. To demonstrate this, in Figs.~2 and 3 we plot $\Phi(n_1, n_2(n_1))$ upon increasing $\mu_{1}$ and thus  evolving the system across the putative transition to the BEC state along two paths in the exemplary putative diagram of Fig.~\ref{Fig_1}. The two paths correspond to varying $\mu_1$ at fixed $\mu_2$ such that $\overline{\mu}_{2}=\overline{\mu}_2^{(1,c)}(\overline{\mu}_{1}=0.43)\approx 0.134$ and $\overline{\mu}_2=\overline{\mu}_2^{(1,c)}(\overline{\mu}_{1}=0.50)\approx 0.445$; for the definitions of dimensionless quantities see Appendix 2.  In both cases the blue line is crossed below the intersection points of the blue dashed and red dashed lines which, for the parameters values chosen for the plot in Fig. 1, is $\overline{\mu}_{1} = \overline{\mu}_{2} \approx 0.522$. 
\begin{figure}
\includegraphics[width=8.5cm]{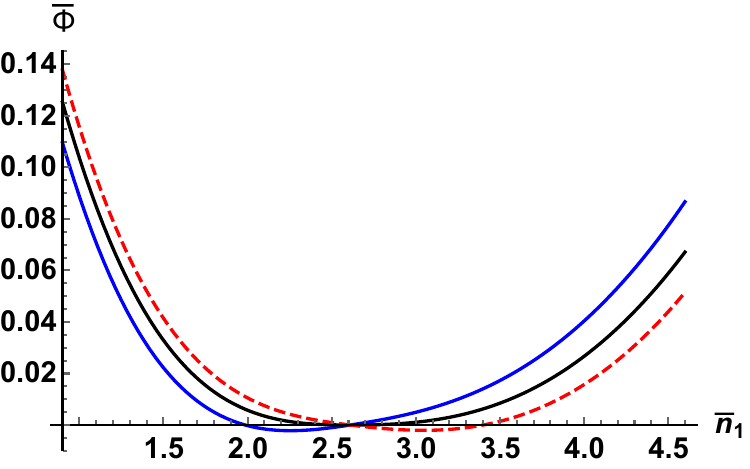}
\caption{Plot of $\overline{\Phi}(\overline{n}_1, \overline{n}_2(\overline{n}_1))$ as the system is tuned through Bose-Einstein  condensation varying $\mu_1$ at fixed $\mu_2$ such that $\overline{\mu}_{2}=\overline{\mu}_2^{(1,c)}(\overline{\mu}_{1}=0.43)\approx 0.134 $ (the remaining parameters as in Fig.~1). The dashed curve corresponds to the state involving the BEC, where $n_1>n_{1,c}$. The minimum of $\overline{\Phi}$ evolves smoothly and at $\overline{\mu}_1=0.43$ is located at $n_1=n_{1,c}$. The system exhibits a second-order transition at the point located on the blue dashed line in Fig.~1. The solid blue curve corresponds to $\overline{\mu}_1=0.42$, the dashed red curve corresponds to $\overline{\mu}_1=0.438$.  The black curve is located at $\overline{\mu}_1=0.43$.  The plot parameters are $\overline{a}_1 = \overline{a}_2=0.1$, $\overline{a}_{12}=0.5$, $\kappa = 1$ (see Appendix 2 for the definitions of the dimensionless variables). } 
\label{Fig_2}
\end{figure} 
\begin{figure}
\includegraphics[width=8.5cm]{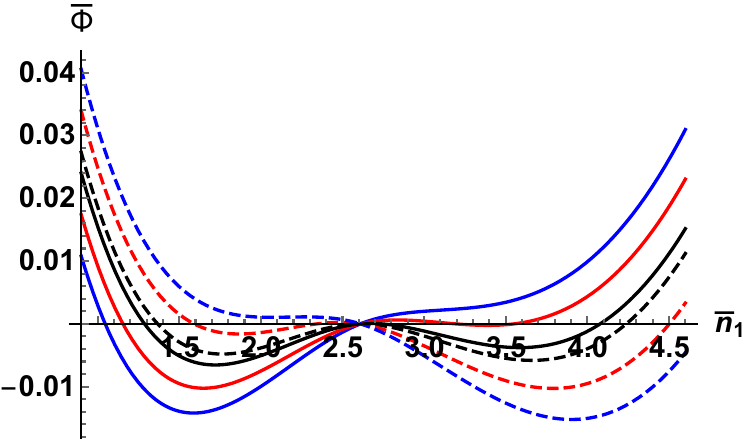}
\caption{Plot of $\overline{\Phi} (\overline{n}_1, \overline{n}_2(\overline{n}_1))$ as the system is tuned through Bose-Einstein condensation varying $\mu_1$ at fixed $\mu_2$ such that $\overline{\mu}_{2}=\overline{\mu}_2^{(1,c)}(\overline{\mu}_{1}=0.50)$ (the remaining parameters as in Fig.~1). The dashed curves correspond to the state involving the BEC, where $n_1>n_{1,c}$. The system exhibits a first-order transition characterized by a jump of $n_1$. The solid blue curve corresponds to $\overline{\mu}_1=0.49$, the dashed blue curve corresponds to $\overline{\mu}_1=0.508$.  The dashed black curve is at $\overline{\mu}_1=0.50$, where $\overline{\Phi}$ exhibits a maximum at $\overline{n}_{1,c}$, marking a point $(\overline{\mu}_1=0.50, \,  \overline{\mu}_2=\overline{\mu}_2^{(1,c)}(0.50)$) located on the blue curve in Fig. 1 with a (false) transition. The true transition, where the global minimum of $\overline{\Phi}$ changes discontinuously  is located slightly to the left from the dashed blue line in Fig.~\ref{Fig_1} (i.e. at a $\overline{\mu}_1<0.50$). The plot parameters are $\overline{a}_1 = \overline{a}_2=0.1$, $\overline{a}_{12}=0.5$, $\kappa = 1$ (see Appendix 2 for the definitions of the dimensionless variables).
 } 
\label{Fig_3}
\end{figure} 

These results clearly demonstrate the necessity of performing careful checks of the nature of the solutions to the saddle-point equations [Eq.~(\ref{n1eq},\ref{n2eq})],  and identifying the ones corresponding to true global minima of $\Phi(n_1, n_2(n_{1}))$ depending on the system parameters. 

We finally observe that the picture of Fig.~1 is qualitatively stable with respect to variation of temperature. For large $T$ one observes scaling of the entire transition line (arising from Eq.~(\ref{mu2ceq}) as $\sim T^{3/2}$, see the following sections for further discussion.   

\section{Phase diagrams}
We now execute the procedure described in Sec.~II and III to project out the phase diagrams. In all of the numerical analysis we restrict to the mass-balanced case $m_1=m_2$. As implied by the structure of the equations, two cases must be distinguished depending on the sign of the quantity $D=a_1 a_2 - a_{12}^2$. The relevance of this parameter was recognized already in earlier literature \cite{Ao_1998, Esry_1997, Schaeybroeck_2013}, where it was found that for $D<0$ the two condensates cannot coexist. We address the distinct  situations separately below.    
\subsection{Case $a_1 a_2- a_{12}^2<0$}
A representative projection of the phase diagram on the $(\overline{\mu}_1, \overline{\mu}_2)$ plane in the case of sufficiently strong interspecies repulsion and low $T$ is given in Fig.~4. From now on we refer to the phase involving condensate of type-1 particles (but not type-2 particles) as the BEC$_1$ phase (and BEC$_2$ analogously). As BEC$_{12}$ we denote a phase where particles of both types form condensates.
We clearly identify a triple point, where the normal and two Bose-Einstein condensed phases BEC$_1$ and BEC$_2$ coexist, as well as two tricritical points, above which the transition between the normal and the BEC$_i$ phases becomes first-order. While the second-order transition lines coincide with the lines $\mu_2^{(1,c)}$, $\mu_2^{(2,c)}$ computed analytically in Sec.~IV [(Eq.~(\ref{mu2ceq})], the first-order transition lines are shifted from them (as is clear from Fig.~3, but is not visible in the scale of Fig.~4). We found no region of stability of the BEC$_{12}$ phase for the present choice of parameter values (see however Sec.~VB). 
\begin{figure}
\includegraphics[width=8.5cm]{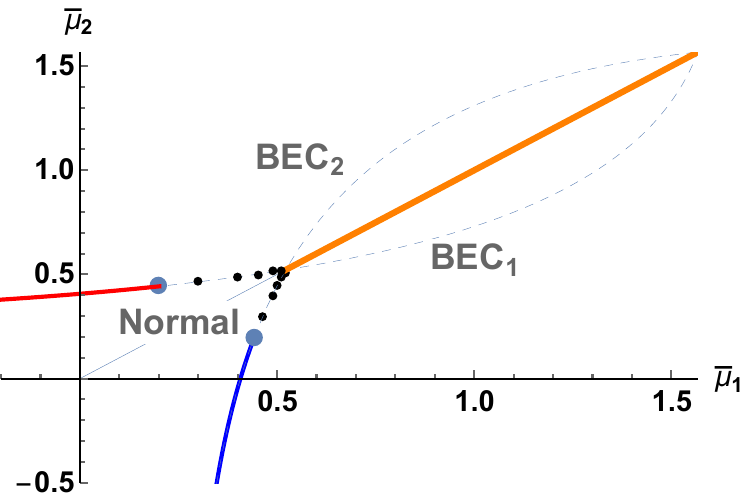}
\caption{Projection of the phase diagram on the $(\mu_1, \mu_2)$ plane for sufficiently strong interspecies coupling $a_{12}$ for $a_1 a_2- a_{12}^2<0$.  The transition between the normal and BEC$_i$ phases is first-order (black points) in the vicinity of the triple point located at $\overline{\mu}_1=\overline{\mu}_2\approx 0.522$ and becomes continuous (solid blue and red lines) below the tricritical points. The transition between the BEC$_1$ and BEC$_2$ phases (solid orange line) is first-order and located at $\overline{\mu}_1=\overline{\mu}_2$. The dashed curves represent the putative transition lines computed in Sec.~IV (see Eq.~\ref{mu2ceq}). While the second-order phase boundaries coincide with these lines, the first-order transition is shifted away from it - compare Fig.~3. This is not visible in the plot scale.  The plot parameters are $\overline{a}_1 = \overline{a}_2=0.1$, $\overline{a}_{12}=0.5$, $\kappa = 1$ (see Appendix 2 for the definitions of the dimensionless variables).} 
\label{Fig_4}
\end{figure}   
An exemplary plot of the dimensionless densities $\overline{n}_i$ as function of $\overline{\mu}_1$ at fixed $\overline{\mu}_2$ somewhat below the triple point is presented in Fig.~5. The densities change discontinuously at the two first-order transitions between the normal and BEC$_i$ phases. At lower values of $\overline{\mu}_2$ (below the corresponding tricritical points) the densities evolve continuously across Bose-Einstein condensation, while for $\overline{\mu}_2$ above the triple-point value there is a single transition, where both $\overline{n}_1$ and $\overline{n}_2$ change discontinuously when crossing the transition line.    
\begin{figure}
\includegraphics[width=8.5cm]{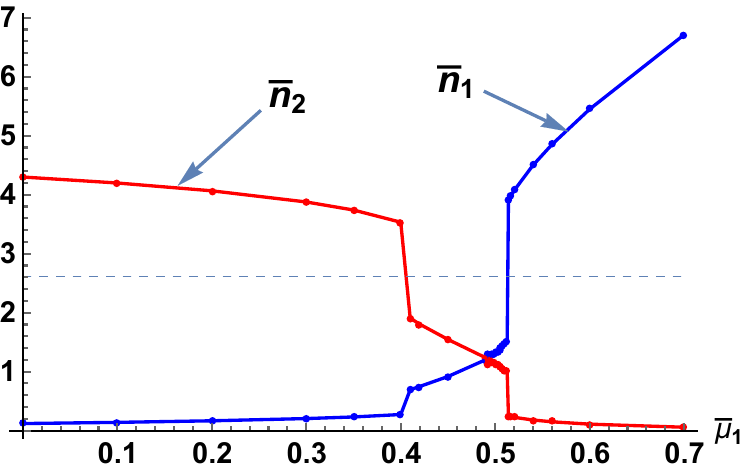}
\caption{Evolution of the densities $\overline{n}_1$ and $\overline{n}_2$ upon varying $\overline{\mu}_1$ at fixed $\overline{\mu}_2=0.49$ and the remaining parameters as in Fig.~4. The system undergoes two first-order phase transitions between the normal and the Bose-Einstein condensed phases. The thin dashed line marks the value $n_{1,c}= n_{2,c}$.  The plot parameters are $\overline{a}_1 = \overline{a}_2=0.1$, $\overline{a}_{12}=0.5$, $\kappa = 1$ (see Appendix 2 for the definitions of the dimensionless variables).
} 
\label{Fig_5}
\end{figure}   

We finally discuss the evolution of the phase diagram depicted in Fig.~4 upon varying temperature. Our analysis indicates that the triple point is present for all values of $T$. On the other hand, the relative distance between the triple and the tricritical points shrinks upon raising $T$. At a certain temperature $T_{\textrm{coll}}$ these points collide, and, for $T>T_{\textrm{coll}}$ the transition between the normal and BEC$_i$ phases is continuous. We plot an exemplary dependence of $\overline{\mu}_1^{\textrm{triple}}$ and $\overline{\mu}_1^{\textrm{tric}}$ on $T$ in Fig.~6 to demonstrate this collision.  
\begin{figure}
\includegraphics[width=8.5cm]{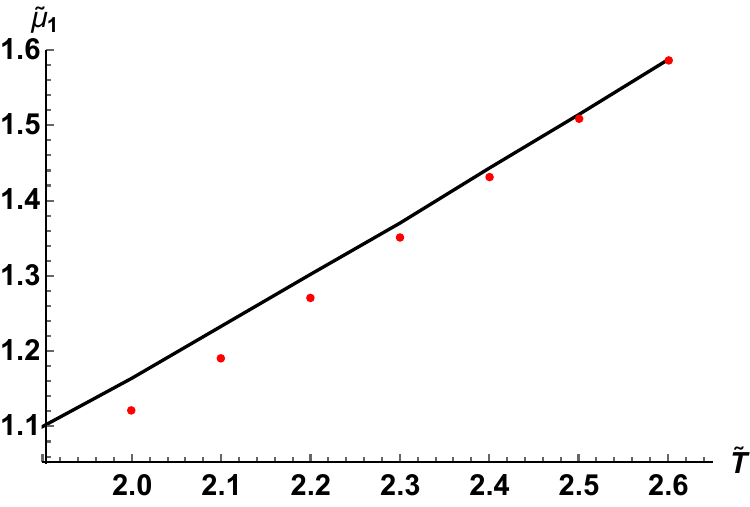}
\caption{The temperature dependence of the location of the triple (black line) and tricritical (red points) points in the $(\mu_1, \mu_2)$ phase diagram. The dimensionless temperature $\tilde{T}$ and chemical potential $\tilde{\mu}_1$ are defined as $\tilde{T}=k_B T [4(2\pi m_1)^3 a_{12}^2]/h^6 =4 {\overline{a}}_{12}^2/\kappa^4$ and $\tilde{\mu}_1 = \mu_1 [4(2\pi m_1)^3 a_{12}^2]/h^6= 4 {\overline{\mu}}_{1}{\overline{a}}_{12}^2/\kappa^4  $. 
The dimensionless parameters are $\frac{\overline{a}_{1}}{ \overline{a}_{12}} =  \frac{\overline{a}_{2}}{ \overline{a}_{12}}=0.2, \kappa=1$. } 
\label{Fig_6}
\end{figure}   

We emphasize that our entire analysis is performed in the {\it grand canonical} ensemble. In this language, occurrence of a first-order transition implies coexistence of two phases characterized by different densities. This is clearly visible in Fig.~3 where this effect is signalled by the occurrence of two degenerate minima of $\bar{\Phi}$. Analogously, when crossing the transition between the BEC$_1$ and BEC$_{2}$ phases (see Fig.~4) at $\bar{\mu}_1=\bar{\mu}_2$, one encounters the coexistence between two thermodynamic states involving condensates. This corresponds to phase separation (or demixing). As will become clear in Sec.~VB, upon manipulating the interaction coupling such that $a_1 a_2-a_{12}$ crosses zero, the system undergoes a mixing transition (well recognized in previous literature), where the two BECs are no longer phase-separated, and no first-order transitions are observed in the phase diagram plotted in the natural variables of the grand-canonical enseble.      

\subsection{Case $a_1 a_2- a_{12}^2>0$}
A significantly different situation occurs for weaker interspecies interaction couplings such that $a_1 a_2- a_{12}^2>0$. For this case we find the transition between the normal and BEC$_i$ phases to be continuous. In addition, we identify a thermodynamic phase BEC$_{12}$ involving condensates of both types of particles. A representative plot is given in Fig.~7. 
\begin{figure}
\includegraphics[width=8.5cm]{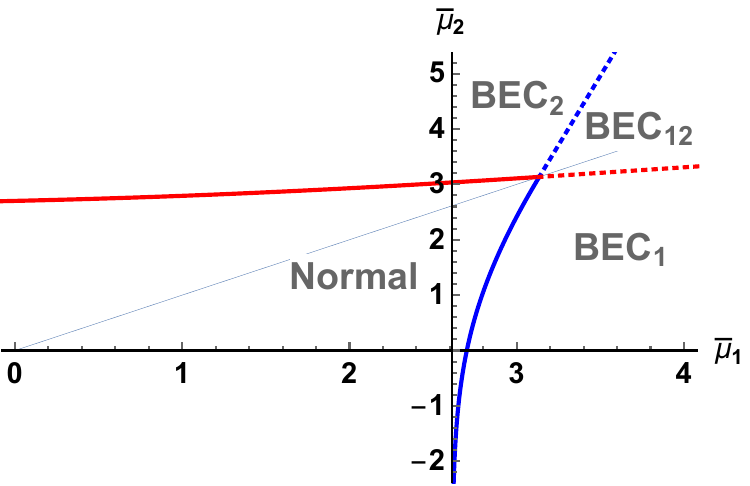}
\caption{A representative projection of the phase diagram on the $(\overline{\mu}_1, \overline{\mu}_2)$ plane for sufficiently weak interspecies coupling (for $a_1 a_2- a_{12}^2>0$). All the involved phase transitions are continuous.   The plot parameters are $\overline{a}_1 = \overline{a}_2=1$, $\overline{a}_{12}=0.2$, $\kappa =1$ (see Appendix 2 for the definitions of the dimensionless variables).}
\label{Fig_7}
\end{figure}   
Remarkably, all the four phases meet at a quadriple point and all the involved transitions are continuous, at least for the ranges of parameters we investigated. An exemplary illustration is presented in Fig.~8, where we plot evolution of the densities $n_1$ and $n_2$ upon varying $\mu_1$ along a horizontal trajectory in Fig.~7. 
\begin{figure}
\includegraphics[width=8.5cm]{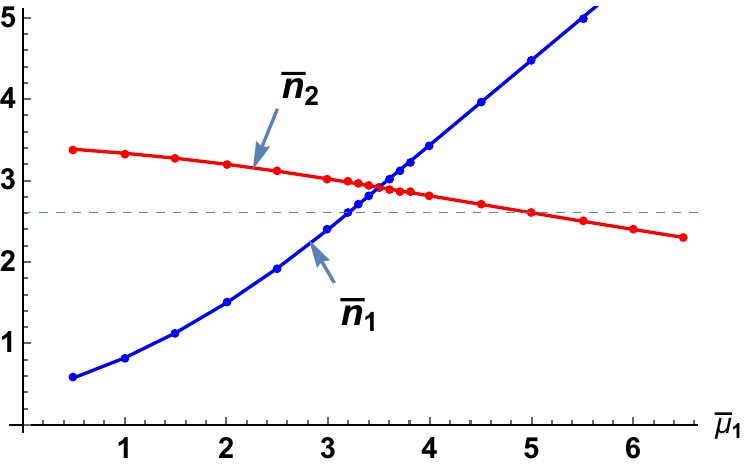}
\caption{Evolution of the densities $\overline{n}_1$ and $\overline{n}_2$ upon varying $\overline{\mu}_1$ at fixed $\overline{\mu}_2=3.5$ and the remaining parameters as in Fig.~7. The system undergoes two continuous phase transitions between the BEC$_2$ and BEC$_{12}$  as well as BEC$_{12}$ and BEC$_{1}$ phases. The thin dashed line marks the value  $n_{1,c}=n_{2,c}$, the transitions take place at the points of intersection between this line and the density curves. 
} 
\label{Fig_8}
\end{figure}   
Using this (numerical) fact as an input, in addition to the previously discussed shape of the normal-BEC$_i$ phase transition line [see Eq.~(\ref{mu2ceq})], one may straightforwardly derive an analytical expression describing the shape of the other phase boundaries (between the BEC$_1$ and BEC$_{12}$ as well as BEC$_2$ and BEC$_{12}$ phases). Focusing on the BEC$_2$-BEC$_{12}$ transition, from Eq.~(\ref{n1eq}) and (\ref{n2eq}), we have 
\begin{align} 
\mu_2-a_2 n_2-a_{12}n_1 = 0 \\
\mu_1-a_1 n_1-a_{12}n_2 = 0\\
n_1 = n_{1,c}\;. 
\end{align}
By eliminating $n_2$ we find the BEC$_2$-BEC$_{12}$ phase boundary given as: 
\begin{equation}
\mu_2(\mu_1)= \frac{a_2}{a_{12}}\mu_1-\frac{n_{1,c}}{a_{12}}\left(a_{1}a_{2}-a_{12}^2\right)\;,   
\end{equation}
which represents a straight line in the $(\mu_1, \mu_2)$ phase diagram. Curiously, the expression is independent of $\lambda_2$, and therefore insensitive to varying the mass of type-2 particles. The line slope is fully controlled by $\frac{a_2}{a_{12}}$ and the entire temperature dependence is in the free term, which is $\sim \lambda_1^{-3}\sim T^{3/2}$ and completely drops out for $\left(a_{1}a_{2}-a_{12}^2\right)\rightarrow 0$. 

It is instructive to also write down the corresponding expression for the BEC$_1$-BEC$_{12}$ phase boundary  
\begin{equation}
\mu_2(\mu_1)= \frac{a_{12}}{a_{1}}\mu_1+\frac{n_{2,c}}{ a_{1}}\left(a_{1}a_{2}-a_{12}^2\right)    
\end{equation}
and investigate the fate of the two lines in the limit $(a_1 a_2- a_{12}^2)\rightarrow 0$, where both of them coincide and are described by
\begin{equation}
\mu_2 (\mu_1)\rightarrow \sqrt{\frac{a_2}{a_1}}\mu_1\;.    
\label{special_case}
\end{equation}
In consequence, upon tuning the interactions towards $(a_1 a_2- a_{12}^2)\rightarrow 0$ the wedge of stability of the BEC$_{12}$ phase in the phase diagram (compare Fig.~7) becomes increasingly acute. In the boundary case $(a_1 a_2- a_{12}^2)= 0$ the BEC$_{12}$ phase is completely expelled from the phase diagram and when slightly modifying the interactions in such a way that $a_1a_2-a_{12}^2<0$ it immediately becomes first order (see Sec. VA). Since achieving this situation requires tuning  two parameters (for example $a_{12}$ and $T$) in the multidimensional parameter space of the system, we expect  the transition between the BEC$_1$ and BEC$_2$ phases to be of a tricritical character. This is in contrast to the case $a_1a_2-a_{12}^2>0$, where the transition can be achieved by tuning only one parameter (e.g. $T$ at fixed interaction couplings and densities).  Note also the lack of any dependence of Eq.~(\ref{special_case}) on the particle masses and temperature. 

We close this section with a comment regarding the relation between the values of the model parameters adopted in the numerical analysis above and those relevant to experimental setups such as ultracold gases of $^7$Li atoms. 
Assuming that the interaction potential has a characteristic strength amplitude $v_0$ and range $r_0$, we  estimate the magnitude of the interaction couplings in our model as $a_{i,j}\approx a=\int_{\mathbb{R}^d}d\vec{x} v(\vec{x})\approx (4/3)\pi r_0^3v_0$. Taking\cite{Ultracold_exper} $(4/3)\pi r_0^3\approx 10\AA$ and $v_0\approx 0.1 \textrm{eV}$ yields $a\approx 1 \textrm{eV} \AA^3$. The scattering length of the Kac model is on the other hand given by $a_s=a/(4a_0)$ (where $a_0=h^2/(2\pi m)\approx 4*10^{-3} \textrm{eV} \AA^2$ for $^7$Li). This leads to $a_s\approx 10^2\AA$. For realistic values\cite{Ultracold_exper} of the cold-atom densities $\rho\approx 10^{-11}\AA^{-3}$ we find $\rho a_s^3\approx 10^{-5}\ll 1$. This indicates that the Lee-Huang-Yang correction to the interaction energy should not be expected to play an important role. We also  observe that the considered order of magnitude of the dimensionless densities $\rho\lambda^3\approx 1$ corresponds to $T\approx 10^{-6}K$, which is in the range of experimentally reasonable values. The value of the dimensionless coupling $\bar{a}=0.1$ (compare the phase diagram of Fig.~4) also leads to $T\approx 10^{-6}K$. 

\section{Liquid-gas type transition in the normal state}
We finally analyze a separate aspect of the system concerning the non-condensed state for sufficiently strong interspecies repulsion $a_{12}$. In this regime we numerically detected an additional first order phase transition between  component-1 rich and component-2 rich normal phases. The analyzed setup is analogous to the one of Sec.~VA, but we now consider significantly larger values of $a_{12}$ (as compared to $a_1$ and $a_2$). The additional transition line extends from the triple line in the ($\mu_1$, $\mu_2$, $\beta$) phase diagram and terminates with a line of critical points (in the three-dimensional parameter space spanned by $\mu_1$, $\mu_2$, and $\beta$). The distance between the triple and the critical lines is controlled by $a_{12}$. In Fig.~9 we plot the function $\Phi(n_1,n_2(n_1))$ at exemplary points on the detected coexistence line, demonstrating the occurrence of the two minima, which indicates coexistence of two phases characterized by different density compositions and involving no condensates. In Fig.~10 we present a projection of the transition surface on the $(\mu_1, \mu_2)$ plane.  
\begin{figure}
\includegraphics[width=8.5cm]{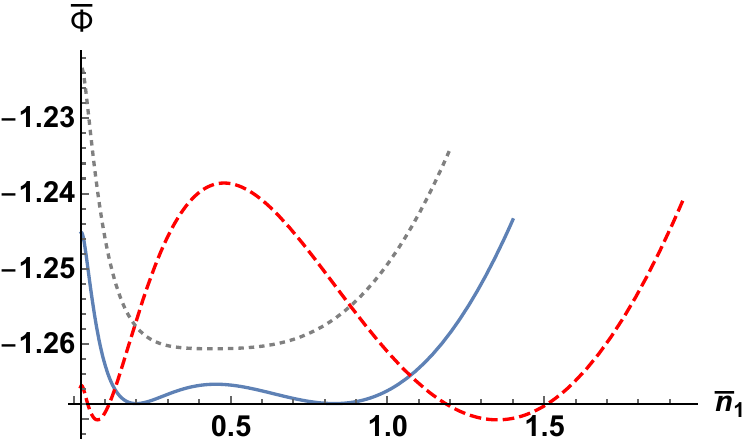}
\caption{Plot of $\overline{\Phi}(\overline{n}_1,\overline{n}_2(\overline{n}_1))$ in the normal phase for a sequence of values of $\overline{\mu}_1=\overline{\mu}_2$ in the regime of large interspecies coupling $\overline{a}_{12}$. The dotted line corresponds to $\overline{\mu}_1=\overline{\mu}_2=-0.05$, the solid line to $\overline{\mu}_1=\overline{\mu}_2=0$, and the dashed line to $\overline{\mu}_1=\overline{\mu}_1=0.1$. The system exhibits two coexisting phases for $\mu_1=\mu_2$ sufficiently large. 
The plot parameters are $\overline{a}_1 = \overline{a}_2=0.1$, $\overline{a}_{12}=2$, $\kappa=1$.  The curves were shifted vertically for better clarity of the illustration. Upon slightly modifying one of the chemical potentials, the degeneracy is removed and one of the states becomes metastable.  See also Fig.~10. 
} 
\label{Fig_9}
\end{figure} 
\begin{figure}
\includegraphics[width=8.5cm]{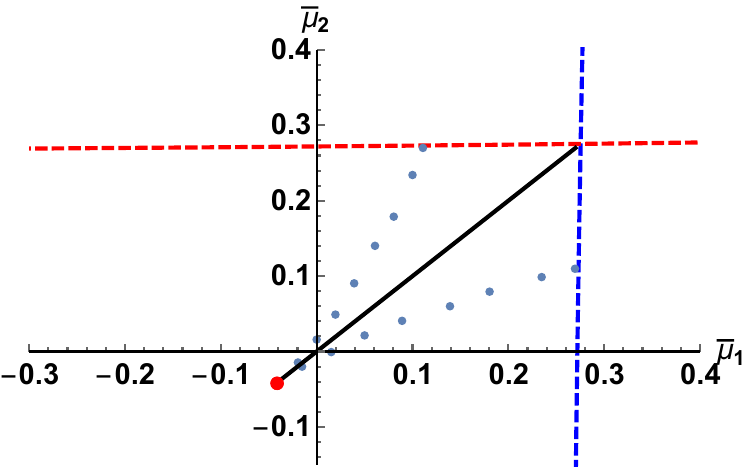}
\caption{Projection of a portion of the phase diagram on the ($\overline{\mu}_1, \overline{\mu}_2$) plane in the regime of large interspecies couplings $\overline{a}_{12}$. The solid diagonal line denotes a first-order transition between two (component-1 rich and component-2 rich) normal phases. The line terminates with a critical point (the red dot). The light blue points represent the spinodal lines marking the region, where metastable states exist. The dashed lines demarkate the regions, where the condensates become present and were determined according to the procedure described in Sec.~IV. and illustrated in Fig.~1. The plot parameters are $\overline{a}_1 = \overline{a}_2=0.1$, $\overline{a}_{12}=2$, $\kappa=1$, compare  also Fig.~9.
} 
\label{Fig_10}
\end{figure}

The transition is reminiscent of those widely considered in the context of classical mixtures and also bears similarity to the classical liquid-gas transition. Strikingly, it does not require the occurrence of any interparticle interactions with attractive components. We emphasize that the presence of an attractive tail in the interaction potential in {\it classical} fluids  in indispensable for the occurrence of van der Waals type transitions. Our present result indicates that in case of Bose systems the role of such interaction may be taken over by quantum statistics. 

Our present study of this aspect of the Bose mixture is unfortunately restricted to numerical analysis. Our results demonstrate a generic occurrence of the liquid-gas type transition provided $a_{12}$ is sufficiently large. At this point we are not able to address the natural question concerning the scale determining the onset of this transition. We cannot rule out the possibility that it in fact always  occurs provided $a_1 a_2-a_{12}^2<0$, but  for $a_1 a_2-a_{12}^2$ close to zero is present only in a tiny region of the phase diagram, which is hard to resolve numerically. This obviously calls for further clarifying studies. 
   
\section{Conclusion}
Bose-Einstein condensation is commonly recognized to be a generically continuous phase transition and its realization as a first-order transition poses an interesting problem both from theoretical and experimental perspectives. In this paper, using an exactly soluble, mean-field type model, we have demonstrated such a possibility in a very simple setup involving a Bose mixture with purely repulsive interactions. We have shown that in the mass-balanced case, for sufficiently strong interspecies interactions, which fulfill the condition $a_1 a_2- a_{12}^2<0$,  Bose-Einstein condensation is realized as a first-order transition in the vicinity of the triple point. We have demonstrated a structural change of the phase diagram  occurring at $a_1 a_2- a_{12}^2=0$ [see Fig.~4 and Fig.~7], where the phase diagram viewed in the $(\mu_1, \mu_2, T)$ space features a two-dimensional surface of tricritical points. In addition, for sufficiently strong (repulsive) interspecies coupling, we identified an additional first-order phase transition between component-1 rich and component-2 rich normal phases, which does not seem to have been discussed in  literature. Our predictions are certainly open to verification via experiments and numerical simulations. This concerns both the 1-st order character of the BEC transition in the vicinity of the triple point, as well as the existence of the additional transition within the region of the phase diagram hosting the normal phase. Even though our study relies on a model characterized by long-ranged interparticle interactions, its predictions are closely related (and in some aspects equivalent) to those of the Hartree-Fock treatment of the dilute Bose gases with short-ranged forces. This provides good reasons to believe that our findings are of relevance also to such situations. We observe on the other hand, that long-range interacting
potentials can also be experimentally realized\cite{Landig_2016}.   
  On the theory side there are a number of interesting extensions of the present study, involving in particular systems with mass imbalance, attractive interspecies interactions, as well as beyond mean-field effects,\cite {Ceccarelli_2015, Ceccarelli_2016, Utesov_2018, Ota_2020, Isaule_2021, Isaule_2022, Spada_2023_2} which we relegate to future studies.  

\begin{acknowledgments}
 PJ thanks Maciej Łebek for a useful discussion. MN acknowledges support from the Polish National Science Center via grant2021/43/B/ST3/01223, PJ via grant 2017/26/E/ST3/00211, and KM via grant 2020/37/B/ST2/00486.
\end{acknowledgments}
\section*{Appendix 1}
In this Appendix we derive the expression for the grand canonical partition function of the imperfect Bose mixture, Eq.~(8).  The definition in Eq.~(7) evaluated for the Hamiltonian in Eq.(5) reads 
\begin{align}
\label{bm01}
&\Xi(T,V,\mu_{1},\mu_{2}) = \\
&\sum_{N_{1}=0}^{\infty}\sum_{N_{2}=0}^{\infty} e^{\beta \left(\mu_{1}'N_{1}' +\mu_{2}'N_{2}'- 
\frac{a_{1}'}{2V}N_{1}'^2- \frac{a_{2}'}{2V} N_{2}'^2\right)}Z_{0}^{(1)}(T,V,N_{1})\,Z_{0}^{(2)}(T,V,N_{2})\, ,\nonumber
\end{align}
where $a_{1}'=a_{1}$, $a_{2}' = a_{2} - \frac{a_{12}^2}{a_{1}}$, $\mu_{1}'=\mu_{1}$, $\mu_{2}'=\mu_{2}-\frac{a_{12}}{a_{1}}\mu_{1}$, $N_{1}' = N_{1}+\frac{a_{12}}{a_{1}}N_{2}$, $N_{2}' = N_{2}$, and $Z_{0}^{(i)}(T,V,N_{i})$ denotes the canonical partition function of the ideal Bose gas formed by the $i-$th species. 

First we consider the case $a_{2}'> 0$. In this case we apply twice the identity
\begin{align}
\label{iden01}
\exp\left(-\frac{\gamma_{i}^2}{4\delta_{i}}\right) \,=\, \sqrt{\frac{\delta_{i}}{\pi}} \, \int\limits_{-\infty}^{\infty} dq_{i} \, 
\exp(-\delta_{i} q_{i}^2 + i \gamma_{i} q_{i})\,,
\end{align}
where $\delta_{i} > 0$, and obtain the following expression for the partition function:
\begin{align}
&\Xi(T,V,\mu_{1},\mu_{2}) =   \nonumber \\
&\frac{V\beta}{2\pi \sqrt{a_{1}' a_{2}'}} \sum_{N_{1}=0}^{\infty} \sum_{N_{2}=0}^{\infty}
\int\limits_{-\infty}^{\infty} dq_{1}\int\limits_{-\infty}^{\infty} dq_{2} \,Z_{0}^{(1)}(T,V,N_{1})\,Z_{0}^{(2)}(T,V,N_{2})   \nonumber \\
&  \,e^{\,\,\sum\limits_{i=1}^{2}\left(- \frac{V\beta}{2a_{i}'}q_{i}^2 + iq_{i}\beta\,
\left[N_{i}'-\frac{V}{a_{i}'}\,(\mu_{i}' - \alpha_{i})\right] + \beta \alpha_{i} N_{i}' +\frac{V\beta}{2a_{i}'}(\mu_{i}'-\alpha_{i})^2\right)} =   \nonumber \\
&-\frac{V\beta}{2\pi \sqrt{a_{1}' a_{2}'}} \sum\limits_{N_{1}=0}^{\infty} \sum\limits_{N_{2}=0}^{\infty}
\int\limits_{\alpha_{1}-i\infty}^{\alpha_{1}+i\infty} dt_{1}\int\limits_{\alpha_{2}-i\infty}^{\alpha_{2}+i\infty} dt_{2}\,\, e^{\,\sum\limits_{i=1}^{2}\left[\frac{V\beta}{2a_{i}'}(t_{i}-\mu_{i}')^2+ t_{i}\beta N_{i}'\right]} \nonumber \\
& \;\;\; Z_{0}^{(1)}(T,V,N_{1})\;\,Z_{0}^{(2)}(T,V,N_{2}) \,\,,  
\end{align}
where $\alpha_{i}$ ($i\in\{1,2\}$) are arbitrary constants. After performing the summations over $N_{i}$, using the expression for the grand canonical partition function of an ideal Bose gas 
\begin{align} 
 \Xi_{0}^{(i)}(T,V,s_{i}) = \sum_{N_{i}=0}^{\infty} \,e^{\beta s_{i} N_{i}}\, Z_{0}^{(i)}(T,V,N_{i}) = \exp{\left(\frac{V}{\lambda_{i}^3} \,
 g_{\frac{5}{2}}(e^{\beta s_{i}}) \right)}
\end{align} 
and changing the integration variables we obtain the expressions displayed in Eqs (8) and (9). \\
In the case $a_{2}' < 0$ we proceed analogously except that we additionally use the identity 
\begin{align}
\label{tozs1}
\exp{\left(\frac{\gamma^2}{4\delta}\right)} = \sqrt{\frac{\delta}{\pi}}\, 
\int\limits_{-\infty}^{\infty} dq  \exp{\left(-\,\delta q^2 -\gamma q\right)}  \,. 
\end{align}
Following the steps described above we arrive at the following expression 
\begin{align}
&\Xi(T,V,\mu_{1},\mu_{2}) = \nonumber \\ 
&\frac{V\beta}{2\pi \sqrt{a_{1}' |a_{2}'|}}  \sum_{N_{1}=0}^{\infty}  \sum_{N_{2}=0}^{\infty}
\int\limits_{-\infty}^{\infty} dq_{1}\int\limits_{-\infty}^{\infty} dq_{2} \,Z_{0}^{(1)}(T,V,N_{1})\,\,Z_{0}^{(2)}(T,V,N_{2}) \nonumber \\
&\,e^{- \frac{V\beta}{2a_{1}'}q_{1}^2 + iq_{1}\beta\,
\left(N_{1}'-\frac{V}{a_{1}'}\,(\mu_{1}' - \alpha_{1})\right) + \beta \alpha_{1} N_{1}' +\frac{V\beta}{2a_{1}'}(\mu_{1}'-\alpha_{1})^2 } \nonumber\\
&e^{- \frac{V\beta}{2|a_{2}'|}q_{2}^2 - (q_{2}-\alpha_{2})\beta N_{2}' -\frac{V\beta}{2|a_{2}'|}\,\left((\mu_{2}'-\alpha_{2})^2+2q_{2}(\mu_{2}'-\alpha_{2})\right)} =\nonumber \\
& - \frac{i\,V\beta}{2\pi \sqrt{a_{1}' |a_{2}'|}}  \sum_{N_{1}=0}^{\infty}  \sum_{N_{2}=0}^{\infty}
\int\limits_{\alpha_{1}-i\infty}^{\alpha_{1}+\infty} dt_{1}\int\limits_{-\infty}^{\infty} dt_{2} \,\,e^{\,\sum\limits_{i=1}^{2}\left[\frac{V\beta}{2a_{i}'}(t_{i}-\mu_{i}')^2+ t_{i}\beta N_{i}'\right]} \nonumber \\
& \,\, Z_{0}^{(1)}(T,V,N_{1})\,\,Z_{0}^{(2)}(T,V,N_{2}) \,.
\end{align} 
Analogously to the previous case one obtains again Eqs (8) and (9) with the same expression for 
$\Phi(t_{1},t_{2})$ except that now the integration over variable $t_{2}$ is taken along the real axes. \\

\section*{Appendix 2}
In this Appendix we rewrite Eqs. (\ref{n1eq},\ref{n2eq},\ref{Phieq}) using dimensionless quantities: 
\begin{align}
\overline{\mu}_{i} = \beta \mu_{i} \;,
\overline{n}_{i} = n_{i}\lambda_{i}^3,\; \overline{a}_{i}= \beta a_{i} \lambda_{i}^{-3}\;, \nonumber \\ 
\overline{a}_{12}= \beta a_{12} \lambda_{1}^{-\frac{3}{2}}\lambda_{2}^{-\frac{3}{2}} \;, \; \kappa = \left(\frac{\lambda_1}{\lambda_2}\right)^{\frac{3}{2}} \;.
\end{align}
One obtains
\begin{align}
&\overline{n}_{1} = g_{\frac{3}{2}}\left(e^{\overline{\mu}_{1}-\overline{a}_{1}\overline{n}_{1} -\overline{a}_{12}\overline{n}_{2} \kappa} \right) +\;\frac{\lambda_{1}^3}{V}\;\frac{1}{e^{-\left(\overline{\mu}_{1}-\overline{a}_{1}\overline{n}_{1} -\overline{a}_{12}\overline{n}_{2} \kappa\right)}-1}
\end{align}
\begin{align}
&\overline{n}_{2} = g_{\frac{3}{2}}\left(e^{\overline{\mu}_{2}-\overline{a}_{2}\overline{n}_{2} -\overline{a}_{12}\overline{n}_{1} \kappa^{-1} }\right) +\;\frac{\lambda_{2}^3}{V}\;\frac{1}{e^{-\left(\overline{\mu}_{2}-\overline{a}_{2}\overline{n}_{2} -\overline{a}_{12}\overline{n}_{1} \kappa^{-1}\right)}-1}
\end{align}
and 
\begin{align}
&\overline{\Phi}(\overline{n}_{1},\overline{n}_{2}) = \Phi(n_{1},n_{2}) \,\sqrt{\lambda_{1}^3\lambda_{2}^3} = \nonumber \\
&-\;\frac{1}{2}
    \left[\overline{a}_{1}\overline{n}_{1}^2 \kappa^{-1} + \overline{a}_{2}\overline{n}_{2}^2 \kappa + 2 \overline{a}_{12}\overline{n}_{1}\overline{n}_{2}\right] \nonumber \\ 
    &-\frac{1}{\kappa}\,
    g_{\frac{5}{2}}\left(e^{\overline{\mu}_{1}-\overline{a}_{1}\overline{n}_{1} -\overline{a}_{12}\overline{n}_{2} 
    \kappa}\right) -\;\kappa\,
    g_{\frac{5}{2}}\left(e^{\overline{\mu}_{2}-\overline{a}_{2}\overline{n}_{2} -\overline{a}_{12}\overline{n}_{1} 
    \kappa^{-1}}\right) 
    \;\;+\;\; O\left(\frac{1}{V}\right) \nonumber \;.
\end{align}


%

\end{document}